\documentstyle[12pt]{article}

\begin{document}

\sloppy

\begin{flushright}{UT-737\\ January '96}\end{flushright}

\vskip 1.5 truecm

\centerline{\large{\bf Gaugino Condensation }}
\centerline{\large{\bf in the Early Universe}
\footnote{To be published in Physics Letters B}}
\vskip .75 truecm
\centerline{\bf Tomohiro Matsuda
\footnote{matsuda@danjuro.phys.s.u-tokyo.ac.jp}}
\vskip .4 truecm
\centerline {\it Department of Physics, University of Tokyo}
\centerline {\it Bunkyo-ku, Tokyo 113,Japan}
\vskip 1. truecm

\makeatletter
\@addtoreset{equation}{section}
\def\theequation{\thesection.\arabic{equation}}
\makeatother

\vskip 1. truecm

\begin{abstract}
\hspace*{\parindent}
We examine the process of formation of the gaugino condensation 
within a Nambu-Jona-Lasinio type approach.
We construct an effective Lagrangian description for the gaugino
condensation which include a Weyl compensator superfield whose
vacuum expectation value is related to the gaugino condensation.

\end{abstract}
\newpage

\section{Introduction}
\hspace*{\parindent}
There has recently been considerable attention focused on the study of 
supersymmetric models of elementary particle interactions.
This is especially true in the context of grand unification theories,
where remarkable studies have been done in the hope of solving the
gauge hierarchy problem or unifying the gravitational interaction
in the superstring formalism.
Supersymmetric extension of the gravity(supergravity) seems necessary 
in introducing the soft breaking terms and making the cosmological
constant vanish simultaneously.
In supergravity models, the spontaneous breaking of local supersymmetry
or super-Higgs mechanism may generate soft supersymmetry breaking
terms that allow to fulfill such phenomenological requirements.
However, the super-Higgs mechanism implies the existence of a 
supergravity breaking scale, intermediate between Planck scale($M_{p}$)
and weak scale($M_{W}$).
The intermediate scale is expected to be of O($10^{13}$Gev).
Here we expect that this intermediate scale is implemented by the 
mechanism of gaugino condensation in the hidden sector which couples
to the visible sector by gravitational interaction.
The effective action for 
the gaugino condensation is well studied by many authors\cite{leff},
but we cannot solve the problem of the formation of the gaugino condensation
in the early universe in the context of these effective theories.
Usually there arises a potential barrier and the formation
of the condensation is largely suppressed\cite{goldberg}.

In this paper, we examine the formation of the gaugino condensation in
the weak coupling domain and show that there can be a natural phase 
transition of gaugino condensation in the early universe.

\section{Gaugino condensation in the early universe}
\hspace*{\parindent}  
In the standard superfield formalism of the locally supersymmetric action,
we have:
\begin{eqnarray}
  \label{lag}
  S&=&\frac{-3}{\kappa^{2}}\int d^{8}z E exp\left(-\frac{1}{3}
  \kappa^{2} K_{0} \right)\nonumber\\
  &&+\int d^{8}z{\cal E}\left[W_{0}+\frac{1}{4}f_{0}{\cal WW}\right]
    +h.c.
\end{eqnarray}
where we set $\kappa^{2}=8\pi/M_{p}^{2}$.
In the usual formalism of minimal supergravity, the Weyl rescaling is done 
in terms of component fields.
However, in order to understand the anomalous quantum corrections 
to the classical action, we need a manifest supersymmetric formalism,
in which the Weyl rescaling is also supersymmetric.
It is easy to see that the classical action(\ref{lag}) itself is not
super-Weyl invariant.
However, the lack of the super-Weyl invariance can be  remended
with the help of a chiral superfield $\varphi$(Weyl compensator).

For the classical action (\ref{lag}), the K\"ahler function $K_{0}$, the
superpotential $W_{0}$ and the gauge coupling $f_{0}$ are modified\cite{kap}:
\begin{eqnarray}
  \label{modify}
  K_{0} &\rightarrow& K=K_{0}-6\kappa^{-2}{\sf Re} log\varphi \nonumber\\
  W_{0} &\rightarrow& W=\varphi^{3}W_{0}\nonumber\\
  f_{0} &\rightarrow& f=f_{0}+\xi log\varphi
\end{eqnarray}
$\xi$ is the constant that is decided by the super-Weyl anomaly.

Let us examine the simplest case.
We do not include any chiral matter fields and moduli fields, and
we fix the value of $W_{0}$ and $f_{0}$ as:
\begin{eqnarray}
  \label{fix}
  W_{0}&=& \mu^{3}\nonumber\\
  f_{0}&=& \frac{1}{g^{2}}
\end{eqnarray}
and rescale the field $\varphi$ as:
\begin{equation}
  \tilde{\varphi}=\Lambda \varphi
\end{equation}
Finally we have:
\begin{eqnarray}
  \label{fin}
  K&=&K_{0}-6\kappa^{-2}{\sf Re}
  log\left(\frac{\tilde{\varphi}}{\Lambda}\right)
  \nonumber\\
  W&=&\lambda {\tilde{\varphi}}^{3}\nonumber\\
  f&=&\frac{1}{g^{2}}+\xi log\left(\frac{\tilde{\varphi}}{\Lambda}\right)
\end{eqnarray}
where we set $\lambda\equiv\mu^{3}/\Lambda^{3}$.
(We can set $\mu=\Lambda=M_{p}$)

From the equation of motion for the auxirialy field of the super-Weyl
compensator, we have the relation:
\begin{equation}
  \label{aux0}
  \lambda \tilde{\varphi}^{3}=\frac{1}{6}e^{-\kappa^{2}\frac{K}{2}}\xi
  \lambda^{\alpha}\lambda_{\alpha}
\end{equation}
(Here we rescale the Wely compensator and factor out 
$e^{-\kappa^{2}\frac{K}{2}}$ so that the equations looks lile a 
familiar form.)
The scalar potential is:
\begin{eqnarray}
  \label{pot1}
  V_{0}&=&-3e^{\kappa^{2}K}\kappa^{2}|W|^{2}\nonumber\\
  &=&-3e^{\kappa^{2}K}\kappa^{2}\lambda^{2}|\tilde{\varphi}^{3}|^{2}
\end{eqnarray}
Equation of motion for the auxirialy field(\ref{aux0})
 suggests that the eq.(\ref{pot1}) can be interpreted as a four-fermion
interaction of the gaugino:
\begin{equation}
  \label{4f}
  -\frac{1}{12}\frac{\kappa^{2}}{|{\sf Re} f|^{2}}
  \xi^{2}|\lambda^{\alpha}\lambda_{\alpha}|^{2}
\end{equation}
where the factor of ${\sf Re}f$ appears  because we have
rescaled the gaugino fields to have canonical kinetic terms.
This four-fermion interaction becomes strong as ${\sf Re} f$ reaches 0.
The strong coupling point is:
\begin{equation}
  \label{strong}
  \tilde{\varphi}_{s}=\Lambda e^{-\frac{1}{g^{2}\xi}}
\end{equation}
If the universe starts at very small gaugino condensation,
$\tilde{\varphi}$ rolls down the hill(\ref{pot1}).
Near the origin, the shape of the scalar potential 
is not changed by 1-loop effect
because small value of $\tilde{\varphi}$ suggests small interaction.
As $\tilde{\varphi}$ gets large, 1-loop effect starts
to dominate and lifts the effective potential.
Finally, the rolling stops 
near the strong coupling point $\tilde{\varphi}_s$.
This suggests that the gaugino condensation is naturally
formed and the scale is expected to be
$\sim \Lambda e^{-\frac{1}{g^{2}\xi}}$.

For a second example, we include a dilaton superfield $S$.
Now $f_{0}$ is not a constant and depends on the field $S$:
\begin{equation}
  f_{0}=S
\end{equation}
And the K\"ahler potential for the dilaton superfield is:
\begin{equation}
  K_{0}=-log(S+\overline{S})
\end{equation}
Here we should include the effect of the existence of
the dilaton field in the scalar potential.
The tree level scalar potential is:
\begin{equation}
  V_{0}=h_{S}(G^{-1})^{S}_{S}h^{S}-3e^{\kappa^{2}G}
\end{equation}
here the auxirialy field of S is:
\begin{eqnarray}
  \label{aux}
  h_{S}&=&e^{\kappa\frac{G}{2}}G_{S}+\frac{1}{4}f_{S}
  \lambda^{\alpha}\lambda^{\alpha}\nonumber\\
  &=&\frac{1}{4}e^{\kappa\frac{K}{2}}W\frac{1+6S_{R}\xi^{-1}}{S_{R}}
\end{eqnarray}
where we set $G=K+ln(\frac{1}{4}|W|^{2})$ and $S_{R}=S+\overline{S}$.
The tree level potential can be given in a simple form:
\begin{equation}
  V_{0}=e^{\kappa^{2}K}
  \frac{A}{16}|\tilde{\varphi}^{3}|^{2}
\end{equation}
where
\begin{equation}
  A=\lambda^{2}\kappa^{2}\left[\left(1+\frac{6S_{R}}{\xi}\right)^{2}
  -3\right]
\end{equation}
If $S$ is small and $A$ is negative at the early stage of the universe,
we can expect that the tree level potential has the same characteristics
as the simplest model.
The field $\tilde{\varphi}$ rolls down the hill, and finally reaches
at the strong coupling point.
In this case, there is no problematic potential barrier separating
the weak and strong domain.

If the initial value of $S$ is not so small and $A$ is not negative,
 the situation changes.
There appears a problematic potential barrier which suppresses the 
transition. 
We should also note that the global minimum is not stable for $A>0$.
However, we can be optimistic to expect that the instability of the 
true vacuum is not a problem because at the strong coupling point, 
an effective Lagrangian 
constructed in terms of the confined picture should be much more reliable.
In ref.\cite{ross}, cut-off parameter $\Lambda_{c}$ has introduced 
to avoid this instability.
Ref.\cite{ross} corresponds to a special case of our model.

Finally we will comment on the difference between ref.\cite{goldberg}
and Nambu-Jona-Lasinio like approach.
As is shown in ref.\cite{ross}, our tree level Lagrangian can be cast 
to the same form as the effective superpotential
in ref.\cite{goldberg}, so one may wonder why
the difference discussed above arises.
The crucial difference is that $\varphi$ in eq.(\ref{modify}) is
an auxirialy field so the tree level 
scalar potential related to $F_{\varphi}$
does not exist in our model.
One may also wonder which is wrong.
We cannot find a definite answer to this question, but we can say
that the effective Lagrangian obtained from the confined picture
may not be applied near the origin because such a limit
should correspond to the deconfinement and may be singular.

\section{Conclusion}
\hspace*{\parindent}
We examined the formation of a gaugino condensation in the hidden
sector of the supergravity models within a Nambu-Jona-Lasinio type
approach.

First we considered  a simplest model that contains only one gauge
field and with no dependence on moduli fields.
In this simplest model, we have shown that the phase transition can
naturally occur and 1-loop effect can stabilize the vacuum.

We have also examined a model with a dilaton dependent coupling.
If the initial value of $S$ is small and remains small during the
phase transition, we can expect that the phase 
transition can take place.
If the Hubble constant during inflation lifts the potential for $S$,
small initial value of the dilaton field 
can naturally be realized and the phase transition
can occur.
The main difficulty is the stability of the global vacuum when $A>0$.
We should induce a cut-off scale by hand or merely expect that the 
non-perturbative effects like confinement would stabilize the vacuum.

\section*{Acknowledgment}
\hspace*{\parindent}
We thank K.Fujikawa, T.Hotta and K.Tobe for many helpful discussions.

\end{document}